# Occurrence Typing Modulo Theories

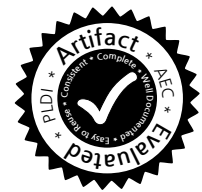

Andrew M. Kent    David Kempe    Sam Tobin-Hochstadt

Indiana University
Bloomington, IN, USA
{andmkent,dkempe,samth}@indiana.edu

## Abstract

We present a new type system combining *occurrence typing*—a technique previously used to type check programs in dynamically-typed languages such as Racket, Clojure, and JavaScript—with *dependent refinement types*. We demonstrate that the addition of refinement types allows the integration of *arbitrary solver-backed reasoning* about logical propositions from external theories. By building on occurrence typing, we can add our enriched type system as a natural extension of Typed Racket, reusing its core while increasing its expressiveness. The result is a well-tested type system with a conservative, decidable core in which types may depend on a small but extensible set of program terms.

In addition to describing our design, we present the following: a formal model and proof of correctness; a strategy for integrating new theories, with specific examples including linear arithmetic and bitvectors; and an evaluation in the context of the full Typed Racket implementation. Specifically, we take safe vector operations as a case study, examining all vector accesses in a 56,000 line corpus of Typed Racket programs. Our system is able to prove that 50% of these are safe with *no new annotations*, and with a few annotations and modifications we capture more than 70%.

*Categories and Subject Descriptors*    F.3.1 [*Specifying and Verifying and Reasoning about Programs*]

*General Terms*    Languages, Design, Verification

*Keywords*    Refinement types, occurrence typing, Racket

## 1. Introduction

Applying a static type discipline to an existing code base written in a dynamically-typed language such as JavaScript, Python, or Racket requires a type system tailored to the idioms of the language. When building gradually–typed systems, designers have focused their attention on type systems with relatively simple goals, e.g. ruling out dynamic type errors such as including a string in an arithmetic computation. These systems—ranging from widely-adopted industrial efforts such as TypeScript [6], Hack [16], and Flow [15] to more academic systems such as Typed Racket [25], Typed Clojure [2], Reticulated Python [29], and Gradualtalk [1]— have been successful in this narrow goal.

However, advanced type systems can express more powerful properties and check more significant invariants than merely the absence of dynamic type errors. Refinement and dependent types, as well as sophisticated encodings in the type systems of languages such as Haskell and ML [11, 30], allow programmers to capture more precise correctness criteria for their programs such as those for balanced binary trees, sized vectors, and much more.

In this paper, we combine these two strands of research, producing a system we dub *Refinement Typed Racket*, or RTR. RTR follows in the tradition of Dependent ML [32] and Liquid Haskell [27] by supporting dependent and refinement types over a limited but extensible expression language. Experience with these languages has already demonstrated that expressive and rich program properties can be captured by a fully-decidable type system.

Furthermore, by building on the existing framework of *occurrence typing*, refinement types prove to be a natural addition to the Typed Racket implementation, formal model, and soundness results. Occurrence typing is designed to reason about dynamic type tests and control flow in existing Racket programs, using propositions about the types of terms and simple rules of logical inference. Extending this logic to encompass refinements of types as well as propositions drawn from solver-backed theories produces an expressive system which scales to real programs. In this paper, we show examples drawn from the theory of linear inequalities and the theory of bitvectors.

```
(: max : [x : Int] [y : Int]
    -> [z : Int #:where (∧ (≥ z x) (≥ z y))])
(define (max x y) (if (> x y) x y))
```
---
**Figure 1.** max with refinement types

---

Figure 1 presents a simple demonstration of integrating refinement types with linear arithmetic. The `max` function takes two integers and returns the larger, but instead of describing it as a simple binary operator on values of type `Int`, as the current Typed Racket implementation specifies, we give a more precise type describing the behavior of `max`.

The syntax for function types in RTR allows for explicit dependencies between the domain and range by giving names to arguments which are in scope in any logical refinements. For the range in this example we use `[z : Int #:where (∧ (≥ z x) (≥ z y))]` as syntactic sugar for a logical refinement on the base type `Int`: `(Refine [z : Int] (∧ (≥ z x) (≥ z y)))`. Note that the `max` function definition does not require any changes to accommodate the stronger type, nor do clients of `max` need to care that the type provides more guarantees than before; the conditional in the body of `max` enables the use of the refinement type in the result, as in most refinement type systems. Typed Racket's pre-existing ability to reason about conditionals means that abstraction and combination of conditional tests properly works in RTR without requiring anything more from solvers for various theories.

With these features we enable programmers to enforce new invariants in existing Typed Racket code and thus evaluate the effectiveness of our type checker on real-world programs. As evidence, we have implemented our system in Typed Racket, including support for linear arithmetic and provably-safe vector access, and automatically analyzed three large libraries totalling more than 56,000 lines of code. We determined that approximately 50% of the vector accesses are provably safe with *no code changes*. We then examined one of the libraries in detail, finding that of the 75% that were not automatically proved safe, an additional 47% could be verified by adding type annotations and minor code modifications.

Our primary contributions are as follows:

1. We present the design of a novel and sound integration between occurrence typing and refinement types drawn from arbitrary logical theories.
2. We describe how to scale our design to a realistic implementation (i.e. Typed Racket).
3. We validate our design by using our implementation to verify the majority of vector accesses in a large Racket library.[1]

---
[1] Our accompanying artifact is available at the following url: https://github.com/andmkent/pldi16-artifact

The remainder of this paper is structured as follows: section 2 reviews the basics of occurrence typing and introduces dependency and refinement types via examples; section 3 formally presents the details of our type system, as well as soundness results for the calculus; section 4 describes the challenges of scaling the calculus to our implementation in Typed Racket; section 5 contains the results of our empirical evaluation of the effectiveness of using refinement types for vector bounds verification; section 6 discusses related work; and section 7 concludes.

## 2. Beyond Occurrence Typing

Occurrence typing—an approach whereby different occurrences of the same identifier may be type checked at different types throughout a program—is the strategy Typed Racket uses to type check idiomatic Racket code [25, 26]. To illustrate, consider a Typed Racket function which accepts either an integer *or* a list of bits as input and returns the least significant bit:

```
(: least-significant-bit :
   (U Int (Listof Bit)) -> Bit)
(define (least-significant-bit n)
  (if (int? n)
      (if (even? n) 0 1)
      (last n)))
```

Type checking the function body begins with the assumption that `n` is of type `(U Int (Listof Bit))` (an *ad hoc* union containing `Int` and `(Listof Bit)`). Typed Racket then verifies the test-expression `(int? n)` is well typed and checks the remaining branches of the program with the following insights:

1. In the then branch we know `(int? n)` produced a non-`#f` value, thus `n` must have type `Int`. The type system can then verify `(if (even? n) 0 1)` is well-typed at `Bit`.
2. In the else branch we know `(int? n)` produced `#f`, implying `n` is not of type `Int`. This fact, combined with our previous knowledge of the type of `n`, tells us `n` must have type `(Listof Bit)`. Thus `(last n)` is also well typed and of type `Bit`.

This strategy of gleaning typed-based information from tests in control flow statements is an essential part of occurrence typing. Instead of only describing the expression `(int? n)` as being of type `Boolean`,

$$\vdash \texttt{(int? n)} : \mathbf{B}$$

RTR adds additional type-based logical information:

$$\vdash \texttt{(int? n)} : (\mathbf{B} \,;\, n \in \mathbf{I} \,|\, n \notin \mathbf{I} \,;\, \emptyset)$$

The first additional element, $n \in \mathbf{I}$, states 'if the result is non-`#f`, then `n` is an integer.' We dub this statement the 'then' proposition, since it is what holds in the first branch of a conditional using `(int? n)` as the test. The second element, $n \notin \mathbf{I}$, tells us 'if the result is `#f`, then `n` is not an integer.' This we dub the 'else' proposition, since it is what holds in the second branch of a conditional using (`int? n`) as the test. The final element, $\emptyset$, tells us the expression is not a term that can be lifted into types.

This additional typed-based logical information and the usual logical connectives (conjunction, disjunction, etc.) are already an integral part of Typed Racket's well-tested type system. By enriching and extending this foundation—adding extensible refinement types and new approaches for reasoning about aliasing and identifiers going out of scope—we can provide a more expressive, theory-extensible system capable of verifying a wider array of practical program invariants.

## 2.1 Occurrence Typing with Linear Arithmetic

Consider how a standard vector access function `vec-ref` might be implemented in a relatively simply-typed language (e.g. standard Typed Racket). In order to ensure we only access valid indices of the vector, our function must conduct a runtime check before performing the raw, unsafe memory access at the user-specified index:

```
(: vec-ref : (∀ {A} (Vecof A) Int -> A))
(define (vec-ref v i)
  (if (≤ 0 i (sub1 (len v)))
      (unsafe-vec-ref v i)
      (error "invalid vector index!")))
```

Although the type for `vec-ref` prevents some runtime errors, invalid indices remain a potential problem. In order to eliminate these, we can extend our new system to consider propositions from the theory of linear integer arithmetic (with a simple implementation of Fourier-Motzkin elimination [7] as a lightweight solver). This allows us to give ≤ a dependent function type where the truth-value of the result reports the intuitively implied linear inequality. We can then design a safe function `safe-vec-ref`:

```
(: safe-vec-ref :
   (∀ {A} [v : (Vecof A)]
         [i : Int #:where (∧ (≤ 0 i)
                             (< i (len v)))]
         -> [res : A]))
(define safe-vec-ref unsafe-vec-ref)
```

Now the type guarantees only provably valid indices are used. While replacing *all* occurrences of `vec-ref` with `safe-vec-ref` in a program may seem desirable, such a change would likely result is programs that no longer type check! One reason for this is the validity of an index is not always apparent at the actual use site. For example, consider a standard vector dot product function:

```
(: safe-dot-prod :
   (Vecof Int) (Vecof Int) -> Int)
(define (safe-dot-prod A B)
  (for/sum ([i (in-range (len A))])
    (* (safe-vec-ref A i)
       (safe-vec-ref B i))))
```

Because there is no explicit knowledge about the length of `B`, our attempt verify one of the indices in `safe-dot-prod` will not type check:

> Type Checker error in (`safe-vec-ref B i`)
> Argument 2, expected:
> (Refine [i : Int] (∧ (≤ 0 i) (< i (len B))))
> but given: Int

In order to type check `safe-dot-prod`, the types for the domain must either be enriched to include the assumption that the vectors are of equal length, or a dynamic check must be added which verifies the assumption at runtime. Also note that without carefully examining the use sites of this function it is difficult to know which solution would be ideal—demanding clients statically verify the property at every call may be an unreasonable requirement. Fortunately a middle ground can be achieved by allowing for both:

```
(: dot-prod :
   (Vecof Int) (Vecof Int) -> Int)
(define (dot-prod A B)
  (unless (= (len A) (len B))
    (error "invalid vector lengths!"))
  (safe-dot-prod A B))
```

Legacy code and clients who cannot easily verify their vectors' lengths may continue to call `dot-prod` while clients wishing to statically eliminate this error may call a safe version which uses a stronger type.

Safe vector access is a simple example of the program invariants expressible with occurrence typing extended with the theory of linear integer arithmetic—we have chosen it for thorough examination because it relates directly to our sizable case study on existing Typed Racket code. Xi [31], however, demonstrates at length in the presentation of Dependent ML how the invariants of far richer programs, such as balanced red-black trees and simple type-preserving evaluators, can be expressed using this same class of refinements.

## 2.2 Occurrence Typing with Bitvectors

Linear arithmetic, however, is merely one example of extending RTR with an external theory. To illustrate, we additionally experimented by adding the theory of bitvectors to RTR. By leveraging Z3's bitvector reasoning [8] we were able to type check the helper function `xtime` found in many implementations of AES [19] encryption. This function computes the result of multiplying the elements of the field $\mathbb{F}_{2^8}$ by $x$ (i.e. polynomials of the form $\mathbb{F}_2[x]/(x^8+x^4+x^3+x+1)$, which AES conveniently represents using a byte):

```
(: xtime : Byte -> Byte)
(define (xtime num)
  (let ([n (AND (* #x02 num) #xff)])
    (cond
      [(= #x00 (AND num #x80)) n]
      [else (XOR n #x1b)])))
```

In this example the type `Byte` is a shorthand for the type
(Refine [b : BitVector] (≤ #x00 b #xff)). To
verify this program, we enrich the types of the relevant bitwise operations (e.g. =, AND, etc.) to include propositions and
refinements relating the values to bitvector-theoretic statements and add bitvector literals to the set of terms which
may be lifted to the type level. Adding the theory of bitvectors and verifying this program proved to be a relatively
straightforward process; in subsection 3.4 we discuss in detail our general strategy for adding new theories to RTR.

## 3. Formal Model

Our base system $\lambda_{RTR}$ is a natural extension of Typed
Racket's previous formal model, $\lambda_{TR}$ [26]; new language
forms and judgments are highlighted.

The typing judgment for $\lambda_{RTR}$ resembles a standard typing judgment except that instead of assigning types, it assigns
*type-results* to well-typed expressions:

$$\Gamma \vdash e : (\tau \,;\, \psi_+ \mid \psi_- \,;\, o)$$

This judgment states that in environment $\Gamma$

- $e$ has type $\tau$;
- if $e$ evaluates to a non-false (i.e. treated as true) value,
  'then proposition' $\psi_+$ holds;
- if $e$ evaluates to false, 'else proposition' $\psi_-$ holds;
- $e$'s value corresponds to the symbolic object $o$.

### 3.1 Syntax

The syntax of terms, types, propositions, and other forms are
given in Figure 2.

**Expressions.** $\lambda_{RTR}$ uses a standard set of expressions
with explicit pair operations for simplicity (so our presentation may omit polymorphism).

**Types.** The universal 'top' type $\top$ is the type which describes all well typed terms. **I** is the type of integers, while **T**
and **F** are the types of the boolean values true and false. Pair
types are written $\tau \times \sigma$. $(\bigcup \vec{\tau})$ describes a 'true' (i.e. untagged) union of its components. For convenience we write
the boolean type $(\bigcup \mathbf{T}\ \mathbf{F})$ as **B** and the uninhabited 'bottom'
type $(\bigcup)$ as $\bot$. Function types consist of a named argument
$x$, a domain type $\tau$, and range type-result $R$ in which $x$ is
bound. $\{x{:}\tau \mid \psi\}$ is a standard refinement type, describing
any value $x$ of type $\tau$ for which proposition $\psi$ holds.

**Propositions.** At our system's core is a propositional logic
with domain specific features. $\mathfrak{tt}$ and $\mathfrak{ff}$ are the trivial and
absurd propositions, while $\wedge$ and $\vee$ represent the conjunction and disjunction of propositions. Type information is expressed by propositions of the form $o \in \tau$ or $o \notin \tau$, which
state that symbolic object $o$ is or is not of type $\tau$ respectively.
$o_1 \equiv o_2$ describes an 'alias' between symbolic objects, stating that the object $o_1$ points to the same value as $o_2$. Finally,
an atomic propositions of the form $\chi^{\mathcal{T}}$ represents a statement
from a theory $\mathcal{T}$ for which $\lambda_{RTR}$ has been provided a sound

| | | |
|---|---|---|
| $n ::=$ | $\ldots -2 \mid -1 \mid 0 \mid 1 \mid 2 \ldots$ | **Integers** |
| $p ::=$ | not $\mid$ add1 $\mid$ int? $\mid \ldots$ | **Primitive Ops** |
| $e ::=$ | | **Expressions** |
| | $\mid x$ | variable |
| | $\mid n \mid$ true $\mid$ false $\mid p$ | base values |
| | $\mid \lambda x{:}\tau.e \mid (e\ e)$ | abstraction, application |
| | $\mid (\textbf{if}\ e\ e\ e)$ | conditional |
| | $\mid (\textbf{let}\ (x\ e)\ e)$ | local binding |
| | $\mid (\textbf{cons}\ e\ e)$ | pair construction |
| | $\mid (\textbf{fst}\ e) \mid (\textbf{snd}\ e)$ | field access |
| $v ::=$ | | **Values** |
| | $\mid n \mid$ true $\mid$ false $\mid p$ | base values |
| | $\mid \langle v, v \rangle \mid [\rho, \lambda x{:}\tau.e]$ | pair, closure |
| $\tau, \sigma ::=$ | | **Types** |
| | $\mid \top$ | universal type |
| | $\mid \mathbf{I} \mid \mathbf{T} \mid \mathbf{F} \mid \tau \times \tau$ | basic types |
| | $\mid (\bigcup \vec{\tau})$ | ad-hoc union type |
| | $\mid x{:}\tau \to R$ | function type |
| | $\mid \{x{:}\tau \mid \psi\}$ | refinement type |
| $\psi ::=$ | | **Propositions** |
| | $\mid \mathfrak{tt} \mid \mathfrak{ff}$ | trivial/absurd prop |
| | $\mid o \in \tau \mid o \notin \tau$ | $o$ is/is not of type $\tau$ |
| | $\mid \psi \wedge \psi \mid \psi \vee \psi$ | compound props |
| | $\mid o \equiv o$ | object aliasing |
| | $\mid \chi^{\mathcal{T}}$ | prop from theory $\mathcal{T}$ |
| $\varphi ::=$ | fst $\mid$ snd | **Fields** |
| $o ::=$ | | **Symbolic Objects** |
| | $\mid \emptyset$ | null object |
| | $\mid x$ | variable reference |
| | $\mid (\varphi\ o)$ | object field reference |
| | $\mid \langle o, o \rangle$ | object pair |
| $R ::=$ | | **Type-Results** |
| | $\mid (\tau \,;\, \psi \mid \psi \,;\, o)$ | type-result |
| | $\mid \exists x{:}\tau.R$ | existential type-result |
| $\Gamma ::=$ | $\overrightarrow{\psi}$ | **Environments** |
| $\rho ::=$ | $\overrightarrow{x \mapsto v}$ | **Runtime Environments** |

**Figure 2.** $\lambda_{RTR}$ Syntax

solver. In this way our logic describes an extensible system
that can be enriched with various theories according to the
needs of the application at hand.

**Fields.** A field allows us to reference a subcomponent of a
structural value. For example, if $p$ is a tree-like structure built
using nested pairs, (fst (snd $p$)) would describe the value
found by accessing the first field of the result of accessing
$p$'s second field. In this model having the fst and snd fields
for pairs suffices; in general, fields for both built-in and user-
defined data types are needed in order to type check real-
world programs. Our vector case study, for example, required
a len field which described a vector's length.

**Symbolic Objects.** Instead of allowing our types to depend on *arbitrary* program terms (as is done in systems with full dependent types), we define a canonical subset of terms called *symbolic objects* which represent the terms which may be lifted to the type level in our system. These objects act as a conservative 'whitelist' of sorts, allowing our type system to work in a full-scale programming language by only considering obviously safe terms (i.e. excluding mutated fields, potentially non-deterministic functions, etc.).

Initially these objects are only used to describe values bound to variables, field accesses, and pairs of objects, while the null symbolic object $\emptyset$ represents terms we do not lift to the type level. These objects (excluding pairs) are what allows standard Typed Racket to type check many dynamic programming idioms. When extending RTR to handle additional theories, the grammar of symbolic objects is extended to include program terms the new theory must reason about.

Finally, when performing standard capture-avoiding substitution we keep symbolic objects in the obvious normal form (e.g. $(\text{fst}\ \langle x, y \rangle)$ is reduced to $x$). Propositions that end up directly referring to $\emptyset$, such as $\emptyset \in \mathbf{I}$, are treated as equivalent to $\mathsf{tt}$ (i.e. meaningless) and are discarded.

**Type-Results.** In order to allow our system to easily reason about more than the just the simple type $\tau$ of an expression, we assign a well typed expression a type-result. In addition to describing an expression's type, a type-result further informs the system by explicitly capturing two additional properties: (1) what is learned when the expression's value is used as the test-expression in a conditional—this is described by the pair of propositions $\psi_+ | \psi_-$ in the type-result—and (2) which symbolic object $o$ the expression's value corresponds to.

Existentially quantified type-results allow types to depend on terms with no in-scope symbolic object. Our usage of existential quantification resembles the technique introduced by Knowles and Flanagan [17] in many ways, except that our usage is restricted to when substitution is simply not possible (i.e. when the variable's assigned expression has a null object).

**Environments.** For simplicity in this model we use an environment built entirely of propositions. In a real implementation it is useful to separate the environment into two portions: a traditional mapping of variables to types along with a set of currently known propositions. The latter can then be used to refine the former during type checking.

**Runtime Environments.** Our runtime environments are standard mappings of variables to closed runtime values, appearing in closures and our big-step reduction semantics.

### 3.2 Typing Rules

The typing judgment is defined in Figure 4 and an executable PLT Redex [9] model is included in our accompanying artifact. The individual rules are those previously used by Typed Racket with only a few minor modifications to incorporate our new forms (i.e. existential type-results and aliases).

T-Int, T-True, T-False, and T-Prim are used for type checking the respective base values, with T-Prim consulting the $\Delta$

$$
\begin{aligned}
\Delta(\text{not}) &= x{:}\top \to (\mathbf{B}\ ;\ x \in \mathbf{F}\ |\ x \notin \mathbf{F}\ ;\ \emptyset) \\
\Delta(\text{add1}) &= x{:}\mathbf{I} \to (\mathbf{I}\ ;\ \mathsf{tt}\ |\ \mathsf{ff}\ ;\ \emptyset) \\
\Delta(\text{int?}) &= x{:}\top \to (\mathbf{B}\ ;\ x \in \mathbf{I}\ |\ x \notin \mathbf{I}\ ;\ \emptyset) \\
\Delta(\text{bool?}) &= x{:}\top \to (\mathbf{B}\ ;\ x \in \mathbf{B}\ |\ x \notin \mathbf{B}\ ;\ \emptyset) \\
\Delta(\text{pair?}) &= x{:}\top \to (\mathbf{B}\ ;\ x \in \top \times \top\ |\ x \notin \top \times \top\ ;\ \emptyset)
\end{aligned}
$$

**Figure 3.** Primitive Types

metafunction described in Figure 3 for primitive operators. Note that the then- and else-propositions are consistent with their being false or non-false. Additionally, by default none of these terms will appear in types and propositions, as signified by the null symbolic object $\emptyset$.

T-Var may assign any type $\tau$ to variable $x$ so long as the system can derive $\Gamma \vdash x \in \tau$. The then- and else-propositions reflect the self evident fact that if $x$ is found to equal false then $x$ is of type $\mathbf{F}$, otherwise $x$ is not of type $\mathbf{F}$. The symbolic object informs the type system that this expression corresponds to the program term $x$.

T-Abs, the rule for checking lambda abstractions, checks the body of the abstraction in the extended environment which maps $x$ to $\tau$. We use the standard convention of choosing fresh names not currently bound when extending $\Gamma$ with new bindings. The type-result from checking the body is then used as the range for the function type, and the then- and else-propositions report the non-falseness of the value.

T-App handles function application, first checking that $e_1$ and $e_2$ are well-typed individually and then ensuring the type of $e_2$ is a subtype of the domain of $e_1$. The overall type-result of the application is the range of the function, $R$, with the symbolic object of the operand, $o_2$, lifted and optionally substituted for $x$. This *lifting substitution* is defined as follows:

$$
\begin{aligned}
R[x \overset{\tau}{\mapsto} \emptyset] &= \exists x{:}\tau.R \\
R[x \overset{\tau}{\mapsto} o] &= R[x \mapsto o]
\end{aligned}
$$

In essence, if the operand corresponds to a value our type system can reason directly about (i.e. its object is non-null), we perform capture avoiding substitution as expected. Otherwise, an existential quantifier à la Knowles and Flanagan [17] is used to capture the argument expression's *precise* type, even though it's exact identity is unknown; this enables the function's range to depend on its argument regardless of whether the term can soundly be lifted to the type level.

T-If is used for conditionals, describing the important process by which information learned from test-expressions is projected into the respective branches. After ensuring $e_1$ is well-typed at some type, we make note of the then- and else-propositions $\psi_{1+}$ and $\psi_{1-}$. We then extend the environment with the appropriate proposition, dependent upon which branch we are checking: $\psi_{1+}$ is assumed while checking the then-branch and $\psi_{1-}$ for the else-branch. The type result of a conditional is simply the type result implied by both branches.

$$
\begin{array}{ll}
\text{T-Int} & \text{T-True} \quad\quad \text{T-False} \quad\quad \text{T-Prim} \\
\Gamma \vdash n : (\mathbf{I}\,;\,\mathsf{tt}\,|\,\mathsf{ff}\,;\,\emptyset) & \Gamma \vdash \mathsf{true} : (\mathbf{T}\,;\,\mathsf{tt}\,|\,\mathsf{ff}\,;\,\emptyset) \quad \Gamma \vdash \mathsf{false} : (\mathbf{F}\,;\,\mathsf{ff}\,|\,\mathsf{tt}\,;\,\emptyset) \quad \Gamma \vdash p : (\Delta(p)\,;\,\mathsf{tt}\,|\,\mathsf{ff}\,;\,\emptyset)
\end{array}
$$

$$
\begin{array}{ccc}
\text{T-Var} & \text{T-Abs} & \text{T-Subsume} \\
\dfrac{\Gamma \vdash x \in \tau}{\Gamma \vdash x : (\tau\,;\,x \notin \mathbf{F}\,|\,x \in \mathbf{F}\,;\,x)} & \dfrac{\Gamma, x \in \tau \vdash e : R}{\Gamma \vdash \lambda x{:}\tau.e : (x{:}\tau \to R\,;\,\mathsf{tt}\,|\,\mathsf{ff}\,;\,\emptyset)} & \dfrac{\Gamma \vdash e : R' \quad \Gamma \vdash R' <: R}{\Gamma \vdash e : R}
\end{array}
$$

$$
\begin{array}{cc}
\text{T-If} & \text{T-Let} \\
\dfrac{\begin{array}{c}\Gamma \vdash e_1 : (\top\,;\,\psi_{1+}\,|\,\psi_{1-}\,;\,\emptyset) \\ \Gamma, \psi_{1+} \vdash e_2 : R \\ \Gamma, \psi_{1-} \vdash e_3 : R\end{array}}{\Gamma \vdash (\mathbf{if}\ e_1\ e_2\ e_3) : R} &
\dfrac{\begin{array}{c}\Gamma \vdash e_1 : (\tau_1\,;\,\psi_{1+}\,|\,\psi_{1-}\,;\,o_1) \\ \psi_x = (x \notin \mathbf{F} \wedge \psi_{x+}) \vee (x \in \mathbf{F} \wedge \psi_{x-}) \\ \Gamma, x \in \tau, x \equiv o_1, \psi_x \vdash e : R_2\end{array}}{\Gamma \vdash (\mathbf{let}\ (x\ e_1)\ e_2) : R_2[x \stackrel{\tau_1}{\longmapsto} o_1]}
\end{array}
$$

$$
\begin{array}{c}
\text{T-App} \\
\dfrac{\Gamma \vdash e_1 : (x{:}\tau \to R\,;\,\mathsf{tt}\,|\,\mathsf{tt}\,;\,\emptyset) \quad \Gamma \vdash e_2 : (\sigma\,;\,\mathsf{tt}\,|\,\mathsf{tt}\,;\,o_2) \quad \Gamma \vdash \sigma <: \tau}{\Gamma \vdash (e_1\ e_2) : R[x \stackrel{\sigma}{\longmapsto} o_2]}
\end{array}
$$

$$
\begin{array}{ccc}
\text{T-Cons} & \text{T-Fst} & \text{T-Snd} \\
\dfrac{\begin{array}{c}\Gamma \vdash e_1 : (\tau_1\,;\,\mathsf{tt}\,|\,\mathsf{tt}\,;\,o_1) \\ \Gamma \vdash e_2 : (\tau_2\,;\,\mathsf{tt}\,|\,\mathsf{tt}\,;\,o_2) \\ R = (\tau_1 \times \tau_2\,;\,\mathsf{tt}\,|\,\mathsf{ff}\,;\,\langle x_1, x_2\rangle)\end{array}}{\Gamma \vdash (\mathbf{cons}\ e_1\ e_2) : R[x_1 \stackrel{\tau_1}{\longmapsto} o_1][x_2 \stackrel{\tau_2}{\longmapsto} o_2]} &
\dfrac{\begin{array}{c}\Gamma \vdash e : (\tau_1 \times \tau_2\,;\,\mathsf{tt}\,|\,\mathsf{tt}\,;\,o) \\ R = (\tau_1\,;\,\mathsf{tt}\,|\,\mathsf{tt}\,;\,(\mathsf{fst}\ x))\end{array}}{\Gamma \vdash (\mathbf{fst}\ e) : R[x \stackrel{\tau_1}{\longmapsto} o]} &
\dfrac{\begin{array}{c}\Gamma \vdash e : (\tau_1 \times \tau_2\,;\,\mathsf{tt}\,|\,\mathsf{tt}\,;\,o) \\ R = (\tau_2\,;\,\mathsf{tt}\,|\,\mathsf{tt}\,;\,(\mathsf{snd}\ x))\end{array}}{\Gamma \vdash (\mathbf{snd}\ e) : R[x \stackrel{\tau_2}{\longmapsto} o]}
\end{array}
$$

**Figure 4.** Typing Judgment

T-Let first checks whether the expression $e_1$ being bound to $x$ is well typed. When checking the body, the environment is extended with the type for $x$, a proposition describing $x$'s then- and else- propositions, and an alias stating that $x$ refers to $o_1$ (i.e. the symbolic object of $e_1$). Since $x$ is unbound outside the body, we perform a lifting substitution of $o_1$ for $x$ on the result as we do with function application.

In order to omit polymorphism we use explicit pair introduction and elimination rules. T-Cons introduces pairs, first checking the types and symbolic objects for $e_1$ and $e_2$. The type-result then includes the product of these individual types, propositions reflecting the non-false nature of the value, and a symbolic pair object (all modulo the two lifting substitutions). Pair elimination forms are checked with T-Fst and T-Snd, which ensure their arguments are indeed a pair before returning the expected type and a symbolic object describing which field was accessed.

### 3.3 Subtyping and Proof System

The subtyping and proof system use a combination of familiar rules from type theory and formal logic.

#### 3.3.1 Subtyping

Figure 5 describes the subtyping relation <: for types, symbolic objects, and type-results.

For objects, the null object $\emptyset$ is the top object and objects are sub-objects of any alias-equivalent object. Pair objects are sub-objects in a pointwise fashion.

All types are subtypes of themselves and of the top type $\top$. A type is a subtype of a union if it is a subtype of any element of the union. Unions are only subtypes of a type if *every* member of the union is a subtype of that type. Function subtyping has the standard contra- and co-variance in the domain and range; in order to reason correctly about dependencies when checking the range, the environment is extended to assign $x$ the more specific domain type. Pair subtyping is standard.

For refinement types we have three rules: S-Weaken states if $\tau$ is a subtype of $\sigma$ in $\Gamma$ then so is any refinement of $\tau$; S-Refine1 and S-Refine2 allow subtyping inquiries about refinements to be translated into their equivalent logical inquiries.

The subtyping relation for type-results relies on subtyping for the type and object, and logical implication for the then- and else-propositions. Since existentially quantified type-results are only used as a tool for type checking, there is only one explicit subtyping rule for them: SR-Exists. This rule resembles the standard existential instantiation rule from first order logic, stating an existentially quantified type-result is a subtype of another type result if the subtyping relation holds in the appropriately extended environment.

#### 3.3.2 Proof System

Figure 6 describes the type-specific portion of the propositional logic for $\lambda_{RTR}$. We omit the introduction and elimination rules for forms from propositional logic, since they are identical to those used by $\lambda_{TR}$ [26] (i.e. resembling those found in any natural deduction system).

L-Sub says an object $o$ is of type $\tau$ when it is a known subtype of $\tau$. L-Not conversely lets us prove object $o$ is not of type $\tau$ when assuming the opposite implies a contradiction. L-Bot serves as an '*ex falso quodlibet*' of sorts, allowing us to draw any conclusion since $\bot$ is uninhabited.

Object aliasing allows us to reason about the statically known equivalences classes of symbolic objects. L-Refl and L-Sym provide reflexivity and symmetry for aliasing, while

## Figure 5. Subtyping

**SO-Equiv**
$$\frac{\Gamma \vdash o_1 \equiv o_2}{\Gamma \vdash o_1 <: o_2}$$

**SO-Null**
$$\Gamma \vdash o <: \emptyset$$

**S-Refl**
$$\Gamma \vdash \tau <: \tau$$

**S-Top**
$$\Gamma \vdash \tau <: \top$$

**S-Union1**
$$\frac{\forall \tau \text{ in } \vec{\tau}.\ \Gamma \vdash \tau <: \sigma}{\Gamma \vdash (\bigcup \vec{\tau}) <: \sigma}$$

**S-Union2**
$$\frac{\exists \sigma \text{ in } \vec{\sigma}.\ \Gamma \vdash \tau <: \sigma}{\Gamma \vdash \tau <: (\bigcup \vec{\sigma})}$$

**SO-Pair**
$$\frac{\Gamma \vdash o_1 <: o_3 \quad \Gamma \vdash o_2 <: o_4}{\Gamma \vdash \langle o_1, o_2 \rangle <: \langle o_3, o_4 \rangle}$$

**S-Pair**
$$\frac{\Gamma \vdash \tau_1 <: \tau_2 \quad \Gamma \vdash \sigma_1 <: \sigma_2}{\Gamma \vdash \tau_1 \times \sigma_1 <: \tau_2 \times \sigma_2}$$

**S-Weaken**
$$\frac{\Gamma \vdash \tau <: \sigma}{\Gamma \vdash \{x{:}\tau \mid \psi\} <: \sigma}$$

**S-Refine1**
$$\frac{\Gamma, x \in \tau, \psi \vdash x \in \sigma}{\Gamma \vdash \{x{:}\tau \mid \psi\} <: \sigma}$$

**S-Refine2**
$$\frac{\Gamma \vdash \tau <: \sigma \quad \Gamma, x \in \tau \vdash \psi}{\Gamma \vdash \tau <: \{x{:}\sigma \mid \psi\}}$$

**S-Fun**
$$\frac{\Gamma \vdash \tau_2 <: \tau_1 \quad \Gamma, x \in \tau_2 \vdash R_1 <: R_2}{\Gamma \vdash x{:}\tau_1 \to R_1 <: x{:}\tau_2 \to R_2}$$

**SR-Result**
$$\frac{\Gamma \vdash \tau_1 <: \tau_2 \quad \Gamma, \psi_{1+} \vdash \psi_{2+} \quad \Gamma \vdash o_1 <: o_2 \quad \Gamma, \psi_{1-} \vdash \psi_{2-}}{\Gamma \vdash (\tau_1 \ ;\ \psi_{1+} \mid \psi_{1-}\ ;\ o_1) <: (\tau_2\ ;\ \psi_{2+} \mid \psi_{2-}\ ;\ o_2)}$$

**SR-Exists**
$$\frac{\Gamma, x \in \tau \vdash R_1 <: R_2}{\Gamma \vdash \exists x{:}\tau.R_1 <: R_2}$$

## Figure 6. RTR-specific Logic Rules

**L-Sub**
$$\frac{\Gamma \vdash o \in \sigma \quad \Gamma \vdash \sigma <: \tau}{\Gamma \vdash o \in \tau}$$

**L-Not**
$$\frac{\Gamma, o \in \tau \vdash \text{ff}}{\Gamma \vdash o \notin \tau}$$

**L-Bot**
$$\frac{\Gamma \vdash o \in \bot}{\Gamma \vdash \psi}$$

**L-Refl**
$$\Gamma \vdash o \equiv o$$

**L-Sym**
$$\frac{\Gamma \vdash o_2 \equiv o_1}{\Gamma \vdash o_1 \equiv o_2}$$

**L-Update+**
$$\frac{\Gamma \vdash o \in \tau \quad \Gamma \vdash (\vec{\varphi}\ o) \in \sigma}{\Gamma \vdash o \in \text{update}^+_\Gamma(\tau, \vec{\varphi}, \sigma)}$$

**L-Update−**
$$\frac{\Gamma \vdash o \in \tau \quad \Gamma \vdash (\vec{\varphi}\ o) \notin \sigma}{\Gamma \vdash o \in \text{update}^-_\Gamma(\tau, \vec{\varphi}, \sigma)}$$

**L-Transport**
$$\frac{\Gamma \vdash \psi(o_1) \quad \Gamma \vdash o_1 \equiv o_2}{\Gamma \vdash \psi(o_2)}$$

**L-Theory**
$$\frac{[\![\Gamma]\!]_\mathcal{T} \Vdash \chi^\mathcal{T}}{\Gamma \vdash \chi^\mathcal{T}}$$

**L-TypeFork**
$$\frac{\Gamma \vdash \langle o_1, o_2 \rangle \in \tau_1 \times \tau_2}{\Gamma \vdash o_1 \in \tau_1 \wedge o_2 \in \tau_2}$$

**L-ObjFork**
$$\frac{\Gamma \vdash \langle o_1, o_2 \rangle \equiv \langle o_3, o_4 \rangle}{\Gamma \vdash o_1 \equiv o_3 \wedge o_2 \equiv o_4}$$

**L-RefI**
$$\frac{\Gamma \vdash o \in \tau \quad \Gamma \vdash \psi[x \mapsto o]}{\Gamma \vdash o \in \{x{:}\tau \mid \psi\}}$$

**L-RefE**
$$\frac{\Gamma \vdash o \in \{x{:}\tau \mid \psi\}}{\Gamma \vdash \psi[x \mapsto o]}$$

L-Transport allows us replace alias-equivalent objects in any derivable proposition (giving us transitivity). L-ObjFork and L-TypeFork provide a means for reducing claims about object pairs to be claims about their fields.

L-Update+ and L-Update– play a key role in our system, allowing positive and negative type statements to refine the known types of objects. Roughly speaking, if we know an object $o$ is of type $\tau$, updating some field $(\varphi_n (\ldots (\varphi_0\ o)))$ within the object (abbreviated $(\vec{\varphi}\ o)$) with additional information computes the following: if we know $(\vec{\varphi}\ o) \in \sigma$—that the field *is* of type $\sigma$—we update that field's type $\tau'$ to be approximately $\tau' \cap \sigma$ (i.e. a conservative 'intersection' of the two types); conversely, updating a field's type $\tau'$ with the knowledge that the field *is not* $\sigma$ updates the field to be approximately $\tau' - \sigma$ (i.e. the 'difference' between the two). A full definition of update is given in Figure 7.

L-RefI and L-RefE construct and eliminate refinement types in the expected ways, essentially saying that the proposition $o \in \{x{:}\tau \mid \psi\}$ is equivalent to the compound proposition $o \in \tau \wedge \psi[x \mapsto o]$.

$$
\begin{aligned}
\text{update}^\pm_\Gamma(\tau_1 \times \tau_2, \vec{\varphi}{::}\text{fst}, \sigma) &= \text{update}^\pm_\Gamma(\tau_1, \vec{\varphi}, \sigma) \times \tau_2 \\
\text{update}^\pm_\Gamma(\tau_1 \times \tau_2, \vec{\varphi}{::}\text{snd}, \sigma) &= \tau_1 \times \text{update}^\pm_\Gamma(\tau_2, \vec{\varphi}, \sigma) \\
\text{update}^+_\Gamma(\tau, \epsilon, \sigma) &= \text{restrict}_\Gamma(\tau, \sigma) \\
\text{update}^-_\Gamma(\tau, \epsilon, \sigma) &= \text{remove}_\Gamma(\tau, \sigma) \\
\text{update}^\pm_\Gamma((\bigcup \vec{\tau}), \vec{\varphi}, \sigma) &= (\bigcup \overrightarrow{\text{update}^\pm_\Gamma(\tau, \vec{\varphi}, \sigma)})
\end{aligned}
$$

$$
\begin{aligned}
\text{restrict}_\Gamma(\tau, \sigma) &= \bot \quad \text{if } \tau \cap \sigma = \emptyset \\
\text{restrict}_\Gamma((\bigcup \vec{\tau}), \sigma) &= (\bigcup \overrightarrow{\text{restrict}_\Gamma(\tau, \sigma)}) \\
\text{restrict}_\Gamma(\{x{:}\tau \mid \psi\}, \sigma) &= \{x{:}\text{restrict}_\Gamma(\tau, \sigma) \mid \psi\} \\
\text{restrict}_\Gamma(\tau, \sigma) &= \tau \quad \text{if } \Gamma \vdash \tau <: \sigma \\
\text{restrict}_\Gamma(\tau, \sigma) &= \sigma \quad \text{otherwise}
\end{aligned}
$$

$$
\begin{aligned}
\text{remove}_\Gamma(\tau, \sigma) &= \bot \quad \text{if } \Gamma \vdash \tau <: \sigma \\
\text{remove}_\Gamma((\bigcup \vec{\tau}), \sigma) &= (\bigcup \overrightarrow{\text{remove}_\Gamma(\tau, \sigma)}) \\
\text{remove}_\Gamma(\{x{:}\tau \mid \psi\}, \sigma) &= \{x{:}\text{remove}_\Gamma(\tau, \sigma) \mid \psi\} \\
\text{remove}_\Gamma(\tau, \sigma) &= \tau \quad \text{otherwise}
\end{aligned}
$$

**Figure 7.** Update metafunction

Finally, a proposition $\chi^{\mathcal{T}}$ from theory $\mathcal{T}$ is derived using L-Theory. This rule consults a solver for theory $\mathcal{T}$ with the relevant knowledge from $\Gamma$.

### 3.4 Integrating Additional Theories

Our system is designed in an extensible fashion, allowing an arbitrary external theory to be added so long as a theory-specific solver is provided. To illustrate, we discuss the linear arithmetic extension we implemented in Typed Racket in order to perform our vector-related case study.

To add this theory, we first must identify the canonical set of program terms which appear in the theory's sentences. For our case study this included integer arithmetic expressions of the form $a_0 x_0 + a_1 x_1 + ... + a_n x_n$ (i.e. linear combinations over $\mathbb{Z}$) and a field which describes a vector's length. We can extend the grammar of fields and symbolic objects to naturally include these terms:

$$\varphi ::= ... \mid \mathsf{len}$$
$$o ::= ... \mid n \mid n \cdot o \mid o + o$$

Now our type system and logic can reason directly about the terms our theory discusses.

We then identify the theory-relevant *predicates* and extend our grammar of propositions to include them:

$$\chi^{LI} ::= o < o \mid o \leqslant o$$

Finally, the types of some language primitives must be enriched so these newly added forms are emitted during type checking. For example, we must modify the typing judgment for integer literals to include the precise symbolic object:

T-Int
$$\Gamma \vdash n : (\mathbf{I} \,;\, \mathsf{tt} \mid \mathsf{ff} \,;\, n)$$

Similarly, primitive functions which perform arithmetic computation, arithmetic comparison, and report a vector's length must be updated to return the appropriate propositions and symbolic objects (similar to how int? and fst are handled in our presentation of $\lambda_{RTR}$).

With these additions in place, a simple function which converts linear integer propositions into solver-compatible assertions allows our system to begin type checking programs with these theory-specific types.

### 3.5 Semantics and Soundness

$\lambda_{RTR}$ uses the big-step reduction semantics described in Figure 8, which notably treats all non-false values as 'true' for the purposes of conditional test-expressions. The evaluation judgment $\rho \vdash e \Downarrow v$ states that in runtime-environment $\rho$, expression $e$ evaluates to the value $v$. A model-theoretic satisfaction relation is used to prove type soundness, just as in prior work on occurrence typing [26].

#### 3.5.1 Models

Because our formalism is described as a type-theory aware logic, it is convenient to examine its soundness using a model-theoretic approach commonly used in proof theory. For $\lambda_{RTR}$ a model is any runtime-value environment $\rho$ and is said to *satisfy* a proposition $\psi$ (written $\rho \models \psi$) when its assignment of values to the free variables of $\psi$ make the proposition a tautology. The details of satisfaction are defined in Figure 8. The satisfaction relation extends to environments in a pointwise manner.

In order to complete our definition of satisfaction, we also require a typing rule for closures:

T-Closure
$$\frac{\exists \Gamma. \; \rho \models \Gamma \qquad \Gamma \vdash \lambda x{:}\tau.e \,:\, R}{\vdash [\rho, \lambda x{:}\tau.e] \,:\, R}$$

The satisfaction rules are mostly straightforward. $\mathsf{tt}$ is always satisfied, while the logical connectives $\vee$ and $\wedge$ are satisfied in the standard ways. Aliases are satisfied when the objects are equivalent values in $\rho$.

The satisfaction rules M-Refine, M-RefineNot1, and M-RefineNot2 allow refinement types to be satisfied by satisfying the type and proposition separately. M-Theory consults a decider for the specific theory in order to satisfy sentences in its domain.

From M-Type we see propositions stating an object $o$ is of type $\tau$ are satisfied when the value of $o$ in $\rho$ is a subtype of $\tau$. Similarly M-TypeNot tells us if an object $o$'s value in $\rho$ has a type which does not overlap with $\tau$, then the proposition $o \notin \tau$ is satisfied.

#### 3.5.2 Soundness

Our first lemma states that our proof theory respects models.

**Lemma 1.** *If $\rho \models \Gamma$ and $\Gamma \vdash \psi$ then $\rho \models \psi$.*

*Proof.* By structural induction on $\Gamma \vdash \psi$ □

With our proof theory and models in sync and our operational semantics defined, we can state and prove the next key lemma for type soundness which deals with evaluation.

**Lemma 2.** *If $\Gamma \vdash e : (\tau \,;\, \psi_+ \mid \psi_- \,;\, o)$, $\rho \models \Gamma$ and $\rho \vdash e \Downarrow v$ then all of the following hold:*

1. *all non-$\emptyset$ structural parts of $o$ are equal in $\rho$ to the corresponding parts of $v$,*
2. *$v \neq \mathsf{false}$ and $\rho \models \psi_+$, or $v = \mathsf{false}$ and $\rho \models \psi_-$, and*
3. *$\Gamma \vdash v : (\tau \,;\, \mathsf{tt} \mid \mathsf{tt} \,;\, \emptyset)$*

*Proof.* By induction on the derivation of $\rho \vdash e \Downarrow v$. □

Now we can state our soundness theorem for $\lambda_{RTR}$.

**Theorem 1.** *(Type Soundness for $\lambda_{RTR}$). If $\vdash e : \tau$ and $\vdash e \Downarrow v$ then $\vdash v : \tau$.*

*Proof.* Corollary of Lemma 2. □

$$
\begin{array}{llllll}
\text{B-Val} & \text{B-Var} & \text{B-Let} & \text{B-Abs} & \text{B-Fst} & \text{B-Snd} \\
\rho \vdash v \Downarrow v & \dfrac{\rho(x)=v}{\rho \vdash x \Downarrow v} & \dfrac{\rho \vdash e_1 \Downarrow v_1 \quad \rho[x:=v_1] \vdash e_2 \Downarrow v}{\rho \vdash (\textbf{let }(x\ e_1)\ e_2) \Downarrow v} & \rho \vdash \lambda x{:}\tau.e \Downarrow [\rho, \lambda x{:}\tau.e] & \dfrac{\rho \vdash e \Downarrow \langle v_1, v_2\rangle}{\rho \vdash (\textsf{fst } e) \Downarrow v_1} & \dfrac{\rho \vdash e \Downarrow \langle v_1, v_2\rangle}{\rho \vdash (\textsf{snd } e) \Downarrow v_2}
\end{array}
$$

$$
\begin{array}{lllll}
\text{B-Beta} & \text{B-Prim} & \text{B-IfTrue} & \text{B-IfFalse} & \text{B-Pair} \\
\dfrac{\rho \vdash e_1 \Downarrow [\rho_c, \lambda x{:}\tau.e] \quad \rho \vdash e_2 \Downarrow v_2 \quad \rho_c[x:=v_2] \vdash e \Downarrow v}{\rho \vdash (e_1\ e_2) \Downarrow v} & \dfrac{\rho \vdash e_1 \Downarrow p \quad \rho \vdash e_2 \Downarrow v_2 \quad \delta(p,v_2)=v}{\rho \vdash (e_1\ e_2) \Downarrow v} & \dfrac{\rho \vdash e_1 \Downarrow v_1 \quad v_1 \neq \textsf{false} \quad \rho \vdash e_2 \Downarrow v}{\rho \vdash (\textbf{if } e_1\ e_2\ e_3) \Downarrow v} & \dfrac{\rho \vdash e_1 \Downarrow \textsf{false} \quad \rho \vdash e_3 \Downarrow v}{\rho \vdash (\textbf{if } e_1\ e_2\ e_3) \Downarrow v} & \dfrac{\rho \vdash e_1 \Downarrow v_1 \quad \rho \vdash e_2 \Downarrow v_2}{\rho \vdash (\textbf{cons } e_1\ e_2) \Downarrow \langle v_1, v_2\rangle}
\end{array}
$$

$$
\begin{array}{llllll}
\text{M-Top} & \text{M-Or} & \text{M-And} & \text{M-Alias} & \text{M-Refine} \\
\rho \models \mathbb{tt} & \dfrac{\rho \models \psi_1 \text{ or } \rho \models \psi_2}{\rho \models \psi_1 \vee \psi_2} & \dfrac{\rho \models \psi_1 \quad \rho \models \psi_2}{\rho \models \psi_1 \wedge \psi_2} & \dfrac{\rho(o_1) = \rho(o_2)}{\rho \models o_1 \equiv o_2} & \dfrac{\rho \models o \in \tau \quad \rho \models \psi[x \mapsto o]}{\rho \models o \in \{x{:}\tau \mid \psi\}}
\end{array}
$$

$$
\begin{array}{lllll}
\text{M-Type} & \text{M-TypeNot} & \text{M-Theory} & \text{M-RefineNot1} & \text{M-RefineNot2} \\
\dfrac{\vdash \rho(o) : \tau}{\rho \models o \in \tau} & \dfrac{\vdash \rho(o) : \sigma \quad \sigma \cap \tau = \varnothing}{\rho \models o \notin \tau} & \dfrac{\llbracket \rho \rrbracket_{\mathcal{T}} \Vdash \chi^{\mathcal{T}}}{\rho \models \chi^{\mathcal{T}}} & \dfrac{\rho \models o \notin \tau}{\rho \models o \notin \{x{:}\tau \mid \psi\}} & \dfrac{\rho \models \neg\psi[x \mapsto o]}{\rho \models o \notin \{x{:}\tau \mid \psi\}}
\end{array}
$$

**Figure 8.** Big-step Reduction and Model Relation

Although this model-theoretic proof technique works quite naturally, it includes the standard drawbacks of big-step soundness proofs, saying nothing about diverging or stuck terms. We could address this by adding an error value of type $\bot$ that is propagated upward during evaluation and modify our soundness claim to show error is not derived from evaluating well-typed terms.

## 4. Scaling to a Real Implementation

Although $\lambda_{RTR}$ describes the essence of our approach, there are additional details to consider when reasoning about a realistic programming language.

### 4.1 Efficient, Algorithmic Subtyping

In order to highlight the essential features of $\lambda_{RTR}$ we chose a more declarative description of the type system. To make this process efficient and algorithmic several additional steps can be taken.

**Hybrid environments.** Instead of working with only a set of propositions while type checking, it is helpful to use an environment with two distinct parts: one which resembles a standard type environment—mapping objects to the currently known positive and negative type information—and another which contains only the set of currently known compound propositions (since all atomic type-propositions can be efficiently stored in the former part). With these pieces in place, it is easy to iteratively refine the standard type environment with the update metafunction while traversing the abstract syntax tree instead of saving *all* logical reasoning for checking non-trivial terms.

**Representative objects.** Another valuable simplification which greatly reduced type checking times was the use of representative members from alias-equivalent classes of objects. By eagerly substituting and using a single representative member in the environment, large complex propositions which conservatively but inefficiently tracked dependencies—such as those arising from local-bindings—can be omitted entirely, resulting in major performance improvements for real world Typed Racket programs.

**Propogating existentials.** Our typing judgments use subsumption to omit the less interesting details of type checking. Making this system algorithmic would not only require the standard inlining of subtyping throughout many of the judgments, but would also require that existential bindings on the type-results of subterms be propagated upward by the current term's type-result. This ensures all identifiers in the raw results of type checking are still bound and frees us from simplifying every intermediate type-result (as our model with subsumption often requires). This technique is thoroughly described in Knowels and Flanagan's [17] algorithmic type system, which served as an important motivation for this aspect of our approach.

### 4.2 Mutation

We soundly support mutation in our type system in a conservative fashion. First, a preliminary pass identifies which variables and fields may be mutated during program execution. The type checker then proceeds to type check the program, omitting symbolic objects for mutable variables and fields. This way, the initial type of a newly introduced variable will be recorded but no potentially unsound assumptions will be made from runtime tests in the code.

An illustrative example of this approach in action was found during our vector access case study and analysis of the Racket `math` library. It contained a module with a variable `cache-size` of type `Int`. The type system ensured any updates to the value of `cache-size` were indeed of type `Int`, but tests on the relative size of the cache—such as (> cache-size n)—failed to produce any logical information about the size of `cache-size`. This failure made it impossible to verify accesses whose correctness relied on the result of this test, since a concurrent thread could easily modify the cache and its size between our testing and performing the operation, invalidating any supposed guarantees. Indeed, without much effort we were able to cause a runtime error in the `math` library by exploiting this fact before patching the offending code.

### 4.3 Type Inference and Polymorphism

Typed Racket (and RTR) relies on local type inference [21] to instantiate type variables for polymorphic functions whenever possible. Since type inference is such an essential part of type checking real programs, we were unable to check any interesting examples until we had accommodated refinement types.

The constraint generation algorithm in local type inference, written $\Gamma \vdash^V_{\bar{X}} S <: T \Rightarrow C$, takes as input an environment $\Gamma$, a set of type variables $V$, a set of unknown type variables $\bar{X}$, and two types $S$ and $T$, and produces a constraint set $C$. Since the implementation of the algorithm already correctly handled when $S$ is a subtype of $T$, we merely needed to add the natural cases which allow constraint generation to properly recurse into the types being refined:

CG-Ref
$$\frac{\Gamma, x \in \tau, \psi_1 \vdash \psi_2 \qquad \Gamma \vdash^V_{\bar{X}} \tau <: \sigma \Rightarrow C}{\Gamma \vdash^V_{\bar{X}} \{x{:}\tau \mid \psi_1\} <: \{x{:}\sigma \mid \psi_2\} \Rightarrow C}$$

CG-RefLower
$$\frac{\Gamma \vdash^V_{\bar{X}} \tau <: \sigma \Rightarrow C}{\Gamma \vdash^V_{\bar{X}} \{x{:}\tau \mid \psi\} <: \sigma \Rightarrow C}$$

CG-RefUpper
$$\frac{\Gamma, x \in \tau \vdash \psi \qquad \Gamma \vdash^V_{\bar{X}} \tau <: \sigma \Rightarrow C}{\Gamma \vdash^V_{\bar{X}} \tau <: \{x{:}\sigma \mid \psi\} \Rightarrow C}$$

This naturally requires maintaining the full environment of propositions throughout the constraint generation process. Although we did not perform a detailed analysis, the annotation burden for polymorphic functions seems unaffected by our changes.

### 4.4 Complex Macros

Racket programmers use a series of `for`-macros for many iteration patterns [10]. This simple dot-product example iterates i from 0 to (sub1 (len A)) to perform the relevant computations:

```
(for/sum ([i (in-range (len A))])
  (* (vec-ref A i)
     (vec-ref B i)))
```

Although initially verifying these vector accesses appears somewhat straightforward, Typed Racket's type checker runs *after* macro expansion. At that point the obvious nature of the original program may be obfuscated in the sea of primitives that emerge, and the system is left to infer types for the newly introduced identifiers and lambda abstractions:

```
(letrec
  ([start 0] [end (len A)]
   [step  1] [initial 0]
   [loop
     (λ (pos acc)
       (cond
         [(< pos end)
          (define i pos)
          (loop (+ step pos)
                (+ acc (* (vec-ref A i)
                          (vec-ref B i))))]
         [else acc]))])
  (loop start initial))
```

Here RTR is left to infer types for both the domain and range of the inner `loop` function (note that its arguments were not even annotatable identifiers in the original program). Initially, our local type inference chooses type `Int` for the position argument `pos`. This might be perfectly acceptable in Typed Racket, since `Int` is a valid argument type for `vec-ref`. However, when attempting to verify the vector access, `Int` is too permissive: it does not express the loop-invariant that `pos` is always non-negative.

In an effort to effectively reason about these macros we experimented with adding an additional heuristic to our inference for anonymous lambda applications: if a variable is, directly or indirectly, used as a vector index within the function, we try the type `Nat` instead of `Int`. This type, combined with the upper-bounds check within the loop, is enough to verify the access in (vec-ref A i) and (vec-ref B i) (assuming they are of equal length). However, the heuristic quickly fails in the reverse iteration case, (in-range (len A) 0 -1) (i.e. where i steps from (sub1 (len A)) to 0) since for the last iteration `pos` is -1 and not a `Nat`.

More advanced techniques for inferring invariants—such as those used by Liquid Types[22]—will be needed if idiomatic patterns such as Racket's `for` are to seamlessly integrate with refinement types.

## 5. Case Study: Safe Vector Access

In order to evaluate RTR's effectiveness on real programs we examined all unique vector accesses[2] in three large libraries written in Typed Racket, totalling more than 56,000 lines of code:

- The `math` library, a Racket standard library covering operations ranging from number theory to linear algebra. It

---

[2] Since we type check programs *after* macro expansion, vector accesses were assessed at this time as well, and accesses in macros were only counted once.

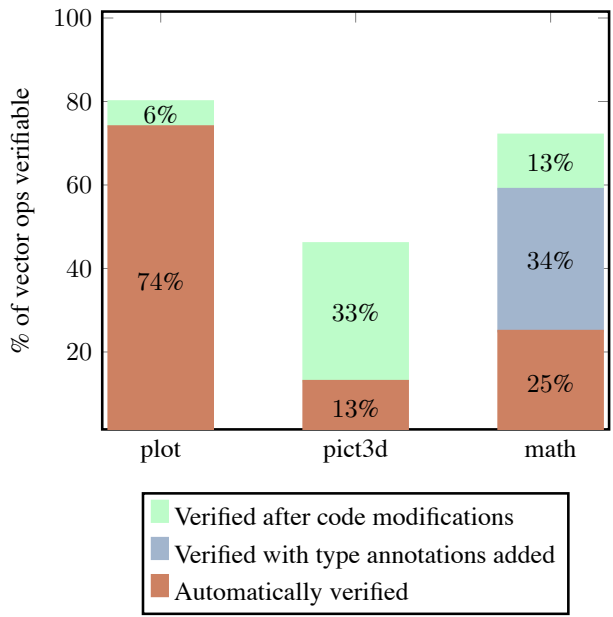

**Figure 9.** `safe-vec-ref` case study

contains 22,503 lines of code and 301 unique vector operations.

- The `plot` library, also a part of Racket's standard library, which supports both 2- and 3-dimensional plotting. It contains 14,987 lines of code and 655 unique vector operations.
- The `pict3d` library,[3] which defines a performant 3D engine with a purely functional interface, has 19,345 lines of code and 129 unique vector operations.

These libraries were chosen because of their size and frequent use of vector operations. During our analysis we tested whether each vector read and write could be replaced with its equivalent `safe-vec-` counterpart and still type check.

To reason statically about vector bounds and linear integer arithmetic we first enriched Typed Racket's base type environment, modifying the type of 36 functions. This included enriching the types of 7 vector operations, 16 arithmetic operations, 12 arithmetic fixnum operations (i.e. operations that work only on fixed-width integers), and the typing of Racket's `equal?`.

We initially verified over 50% of accesses without the aid of additional annotations to the source code. As Figure 9 illustrates, our success rate for entirely automatic verification of vector indices was 74% for `plot`, 13% for `pict3d`, and 25% for `math`. We attribute `plot`'s unusually high automatic success rate relative to the other libraries to a few heavily repeated patterns which are guaranteed to produce safe index-ing: pattern matching on vectors and loops using a vector's length as an explicit bound were extremely common.

---

[3] https://github.com/jeapostrophe/pict3d

For the remaining vector accesses we performed a preliminary review of the `plot` and `pict3d` libraries and an in depth examination of the `math` library.

### 5.1 Enriching the Math Library

For the `math` library we examined each individual access to determine how many of the failing cases our system might handle with reasonable effort. We identified five general categories that describe these initially unverified vector operations:

**Annotations Added.** 34% of the failed accesses were unverified until additional (or more specific) type annotations were added to the original program. In this recursive loop snippet taken from our case study, for example, the `Nat` annotation for the index `i` is not specific enough to verify the vector reference:

```
(let loop ([i : Nat (len ds)] [res : Nat 1])
  (cond
    [(zero? i) res]
    [else
     (loop (- i 1)
           (* res (safe-vec-ref ds i)))]))
```

Using `(Refine [i : Nat] (≤ i (len ds)))` for the type of `i`, however, allows RTR to verify the vector access immediately. As we discussed in subsection 4.4, a more advanced inference algorithm could potentially help by automatically inferring these types. On the other hand, as code documentation these added annotations often made programs easier to understand and helped us navigate our way through the large, unfamiliar code base.

**Code Modified.** 13% of the unverified accesses were verifiable after small local modifications were made to the body of the program. In some cases, these modifications moved the code away from particularly complex macros; other programs presented opportunities for a few well-placed dynamic checks to prove the safety of a series of vector operations. An example of the latter can be seen in the function `vec-swap!`:

```
(: vec-swap! :
   ∀ {A} (Vecof A) Int Int -> Void)
(define (vect-swap! vs i j)
  (unless (= i j)
    (cond
      [(and (< -1 i (len vs))   ;; added
            (< -1 j (len vs)))  ;; added
       (define i-val (safe-vec-ref vs i))
       (define j-val (safe-vec-ref vs j))
       (safe-vec-set! vs i j-val)
       (safe-vec-set! vs j i-val)]
      [else (error "bad index(s)!")])))
```

This function swaps the values at two indices within a vector. Our initial investigation concluded adding constraints to the type was unreasonable for this particular function (i.e.

clients could not easily satisfy the more specific types), however we noticed adding two simple tests on the indices in question allowed us to safely verify four separate vector operations without perturbing any client code. This approach seemed like an advantageous tradeoff in this and other situations and worked well in our experience.

**Beyond our scope.** 22% were unverifiable because, in their current form, their invariants were too complex to describe (i.e. they were outside the scope of our type system and/or linear integer theory). One simple example of this involved determining the maximum dimension `dims` for a list of arrays:

```
(define dims (apply max (map len dss)))
```

Because of the complex higher order nature of these operations, our simple syntactic analysis and linear integer theory was unable to reason about how the integer `dims` related to the vectors in the list `dss`.

**Unimplemented features** 6% of the unverified accesses involved Racket features we had neglected to support during implementation (e.g. dependent record fields), but which seemed otherwise amenable to our verification techniques.

**Unsafe code.** As previously mentioned in subsection 4.2, we discovered 2 vector operations which made unsafe assumptions about a mutable cache whose size could shrink and cause errors at runtime. Both of these correctly did not typecheck using our system and were subsequently patched.

**Total.** In all, 72% of the vector accesses in the math library were verifiable using these approaches without drastically altering any internal algorithms or data representations.[4]

## 6. Related Work

There is a history of using refinements and dependent types to enrich already existing type systems. Dependent ML [32] adds a practical set of dependent types to standard ML to allow for richer specifications and compiler optimizations through simple refinements, using a small custom solver to check constraints. Liquid Haskell [27] extends Haskell's type system with a more general set of refinement types supported by an SMT solver and predicate abstraction. We similarly strive to provide an expressive, practical extension to an existing type system by adding dependent refinements. Our approach, however, seeks to enrich a type system designed *specifically* for dynamically typed languages and therefore is built on a different set of foundational features (e.g. subtyping, 'true' union types, type predicates, etc.).

Some approaches, aiming for more expressive type specifications, have shown how enriching an ML-like typesystem with dependent types and access to theorem proving (automated and manual) provides both expressive and flexible programming tools. ATS, the successor of DML, supports both dependent and linear types as well as a form of interactive theorem proving for more complex obligations [3].

F* [23, 24] adds full dependent types and refinement types (along with other features) to an $F_\omega$-like core while allowing manual and SMT solver-backed discharging of proof obligations. Although our system shares the goal of allowing users to further enrich their typed programs beyond the expressiveness of the core system, we have chosen a simpler, less expressive approach aimed at allowing dynamically typed programs to gradually adopt a simpler set of dependent types.

Chugh et al. [5] explore how extensive use of refinement types and an SMT solver enable type checking for rich dynamically typed languages such as JavaScript [4]. This approach feels similar to ours in terms of features and expressiveness. As seen in our respective metatheories, however, their system requires a much more complicated design and a complex stratified soundness proof; this fact has made it "[difficult to] add extra (basic) typing features to the language" [28]. In contrast, our system uses a well-understood core and does not *require* interaction with an external SMT solver. This allows us to use many common type-theoretic algorithms and techniques—as witnessed by Typed Racket's continued adoption of new features.

Vekris et al. [28] explore how refinements can help reason about complex JavaScript programs utilizing a novel two phase approach. The first phase elaborates the source language into a ML-like target that is checked using standard techniques, at which point the second phase attempts to verify all ill-typed branches are in fact infeasible using refinements in the spirit of Knowles and Flanagan [18] and Rondon et al. [22]. Our single-phase approach, however, does not require elaboration into an ML-like language and allows our system to work more directly with a larger set of types.

Sage's use of a dynamic and static types is similar to our approach for type checking programs. However, their usage of first-class types and arbitrary refinements means their core system is expressive yet undecidable [13]. Our system utilizes a more conservative, decidable core in which only a small set of immutable terms are lifted into types. Because of this, having impure functions and data in the language does not require changes to the type system. Also, our approach only reasons about non-type related theories when they are explicitly added.

Our usage of existential quantification to enable dependent yet abstract reasoning for values no longer in scope strongly resembles the approach described by Knowels and Flanagan [17]. Our design, however, lifts fewer terms into types in general and substitutes terms directly into types when possible. Additionally, our design includes features specifically aimed at dynamic languages instead of refining a more standard type theory.

Ou et al. [20] aim to make the process of working with dependent types more palatable by allowing fine-grained control over the trade-offs between dependent and simple types. This certainly is similar to our system in spirit, but there are several important differences. They choose to automatically

---
[4] Our modified math library can be found in our artifact.

307

insert coercions when dependent fragments and simple types interact, while we do not explicitly distinguish between the two and require explicit code to cast values. Additionally, while they convert their programs from a surface language into an entirely dependently typed language, our programs are translated into dynamically typed Racket code, which is void of any artifacts of our type system. This places us in a more suitable position for supporting sound interoperability between untyped and dependently typed programs.

Manifest contracts [12] are an approach that uses dependent contracts both as a method for ensuring runtime soundness and as a way to provide static typing information. Unlike our system, this method only reasons about explicit casts (i.e. program structure does not inform the type system), and there is no description of how a solver would be utilized to dispatch proof goals.

## 7. Conclusion

Enriching existing programs with stronger static guarantees is the original goal of the scripts-to-programs approach. Its realization in the type system underlying Typed Racket allows Racket programmers to add simple types to their programs with relatively little effort. In this paper, we show how refinement types and an extensible logic allows programmers to continue this process by adding additional invariants to their repertoire which allow for strong new guarantees. Our integration of refinement types with the occurrence typing underlying Typed Racket produces a new system that is both more expressive and simpler than previous approaches.

Additionally, our evaluation demonstrates that despite the relatively simple nature of RTR's dependent types, the invariants that can be expressed are powerful. Our case study of vector operations finds that half of existing operations in a large Typed Racket code base can be already proven safe with many of the remainder checked with simple annotations and changes to the code. We anticipate that other programs, ranging from fixed-width arithmetic to theories of regular expressions [14], can similarly benefit from the strong specifications provided via refinement types.

## Acknowledgments

This work was supported by funding from the National Science Foundation and the National Security Agency. We would like to thank the anonymous reviewers for their helpful feedback on this article, and our colleagues at Indiana University whose helpful discussions influenced our work: Ambrose Bonnaire-Sergeant, David Christiansen, Cameron Swords, Andre Kuhlenschmidt, and Jeremy Siek.